\documentclass[12pt]{iopart}

\usepackage{iopams}
\usepackage{bm}
\usepackage{graphicx}
\usepackage{citesort}

\begin{document}

\title[Ab initio calculations of indium arsenide in the wurtzite phase\dots]{Ab initio calculations of indium arsenide in the wurtzite phase: Structural, electronic and optical properties}

\author{Luis C. O. Dacal}
\address{Instituto de Estudos Avan\c{c}ados, IEAv-CTA, PO Box 6044, 12228-970, S\~{a}o Jos\'{e} dos Campos - SP, Brazil}
\ead{lcodacal@gmail.com}
\author{A. Cantarero}
\address{Materials Science Institute, University of Valencia, PO Box 22085, E46071 Valencia, Spain}
\ead{cantarer@uv.es}
\begin{abstract}
Most of III-V semiconductors which acquires the zincblende phase as bulk materials, adopt the metastable wurtzite phase when grown in form of nanowires. These are new semiconductors, with new optical properties, in particular a different electronic band gap when compared with that grown in the zincblende phase. The electronic gap of wurtzite InAs at the $\Gamma-$point of the Brillouin zone ($E_0$ gap) have been recently measured, $E_0=0.46$ eV at low temperature. The electronic gap at the $A-$point of the Brillouin zone (the equivalent to the $L-$point in the zincblende structure, $E_1$) has also been recently obtained based on a resonant Raman scattering experiment. In this work, we calculate the band structure of InAs in the zincblende and wurtzite phases, using the \textit{full potential linearized augmented plane wave method}, including spin-orbit interaction. The electronic band gap has been improved through the modified Becke-Johnson exchange-correlation potential. Both the $E_0$ and $E_1$ gap agrees very well with the experiment. From the calculations, a crystal field splitting of 0.122 eV and a spin-orbit splitting of 0.312 eV (the experimental value in zincblende InAs is 0.4 eV) has been obtained. Finally, we calculate the dielectric function of InAs in both the zincblende and wurtzite phases and a comparative discussion is given.
\end{abstract}

\submitto{Materials Research Express}
\maketitle

\section{Introduction}
Semiconductor nanowires (NWs) are adaptable building blocks for the next generation of electronic and photonic devices. As emphasized by C. M. Lieber \cite{Lieber2011}, the controlled growth of axial and radial NW heterostructures down to the atomic level will be the underpinning of novel nanoelectronic devices. In the case of InAs NWs, the low effective mass ($0.023\, m_e$) related to its small band gap ($0.35$ eV at room temperature) makes this material particularly interesting in the fabrication of high performance devices (the electron mobility is around 30,000 cm$^2$V$^{-1}$s$^{-1}$ at room temperature). Increasing the In content $x$ in In$_x$Ga$_{1-x}$As-based transistors, a cut-off frequency of 600 GHz can be reached \cite{Dayeh2010}.

As a form of thin layers, only a single monolayer of InAs can be grown on GaAs due to the large lattice mismatch \cite{Li1994}, but in form of NWs, InAs can grow free of strain after the first layers have been formed and the stress has been released, and long NWs of extraordinary crystal quality can be produced \cite{Tizei2009}. The main fields of interest of NWs are energy conversion \cite{Hochbaum2010}, where they can play an important role due to their large surface/volume ratio (the large lateral surface in comparison with a thin film) and the open possibility of fabrication of core/shell structures; quantum computation, based mainly on axial heterostructures in substitution of quantum dots as developing technology; single photon emitters and detectors \cite{Natarajan2012,Hernandez2011}. They are also interesting systems in the area of photonics and quantum cryptography or gas sensors \cite{Chen2013}, taking profit of the large lateral surface of the NWs: a NW ensemble can have a surface up to 1000 times larger as compared with a thin film \cite{Dayeh2010}. In the specific case of InAs, the most common application was as infrared detector, more recently in combination with GaInSb in the form of superlattice \cite{Rogalski2009} with military applications. But a more recent application is in thermoelectricity; the thermal conductivity in InAs NWs is more than one order of magnitude lower than in bulk samples at room temperature \cite{Zhou2011} and much lower than other III-V semiconductors.

Indium arsenide NWs, as other typical III-V materials, show different polytypes when grown in the form of NWs. The most typical polytype is the wurtzite (WZ), or 2H, although it can be also grown in the 4H phase \cite{Kriegner2011}. Most commonly, there is a polytypism mixture of the zincblende (ZB) and WZ phases as has been shown by polarized Raman scattering measurements \cite{Moeller2011}. J. K. Panda  \textit{et al.} \cite{Panda2012} claimed that they can quantify the ZB content within WZ InAs NWs by measuring the temperature dependence of the phonon energies by resonant Raman spectroscopy. A direct gap of 0.46 eV at low temperatures \cite{Moeller2012} and a $E_1$ gap of 2.47 eV at room temperature have been obtained for the WZ phase \cite{Moeller2011a}. In a recent work, I. Zardo \textit{et al.} provided a value of 2.45 eV for the $E_1$ gap, obtained also via resonant Raman scattering measurements \cite{Zardo2013}. The fact that the crystal phase changes from ZB to WZ when going from bulk to NWs can be related to the surface energy or the Coulomb interaction. In general, the electronic and optical properties of InAs NWs are the same as that found in the bulk for the correspondent symmetry. It is necessary to reduce the diameter of the NW to $< 40-50$ nm in order to find electronic quantum confinement \cite{Koblmuller2012} and to 2-3 nm to find phonon confinement \cite{Comas1995}.

Despite of the technological interest on WZ InAs, there are some basic open questions about its physical properties. Probably, the most important one is the value of the InAs band gap in the WZ phase. Some theoretical works predict a higher gap for the WZ phase as compared to that of ZB InAs, but without a satisfactory agreement with the experimental observations \cite{Muranaka,Pryor2010,Zanolli2007}. On the experimental side, there are contradictory values for the WZ InAs band gap. Some works indicate that there is no significant difference in the band gap energy for the WZ and ZB phases of InAs when confinement effects are neglected \cite{Sun2010,Koblmuller2012,Sun2012} while others experiments indicate a higher band gap for the WZ \cite{Tragaardh2007,Moeller2012}.

A few years ago, Zanolli \textit{et al.} \cite{Zanolli2007} calculated the band gap of InAs using the GW approximation neglecting spin-orbit (SO) interaction (there were no experimental data at the time). Actually, they started from the screened-exchange approximation (SXA), using the experimental values of the dielectric constants (the dielectric constant obtained via LDA is too large due to the negative gap problem), and extended the dielectric function to finite frequencies using the generalized plasmon-pole model (PMM). Recently, Larson \textit{et al} have shown that the band gap obtained with the GW correction depends on the kind of PPM used \cite{Larson2013}.  The values obtained for the band gaps in Zanolli's paper \cite{Zanolli2007} were 0.556 and 0.611 eV for the ZB and WZ, respectively. Since SO was not included, they built from the calculated band gap what they called ``experimental gap'' of ZB InAs in a ``scalar relativistic World''. The ``experimental band gaps'' without SO interaction were in good agreement with that calculated with the GW correction.

A few years ago, Becke and Johnson \cite{Becke2006} proposed a simple effective potential to account for the exchange-correlation. Tran and Blaha \cite{Tran2009} modified the potential proposed by Becke and Johnson (mBJ) in order to fit the electronic band gaps of a series of semiconductors, insulators and noble gases. The obtained agreement was of the same order than that obtained with GW methods with a much smaller computational effort. Later, K\"oller \textit{et al.} published a work on the limitations of the method \cite{Koller2011} and one year later they proposed a new parametrization in order to optimize the band gap of a large amount of compounds, 42 in total \cite{Koller2012}. However, in the case of ZB InAs, the re-parametrization lowered the band gap from 0.59 to 0.57 eV, while the experimental band gap at low temperatures is 0.42 eV. This clearly shows the limitation of the new exchange-correlation potential. Nevertheless, the mBJ potential has been successfully used in the past to analyze a quite large number of materials, from half-Heusler topological insulators \cite{Feng2010} to layered semiconductors \cite{Olguin2013}.

The purpose of the present work is to give an accurate description of the WZ InAs band structure. Thus, we have slightly modified the previous parametrization \cite{Tran2009,Koller2012} in order to obtain a realistic band gap for the InAs in the ZB phase. As it is well known, the valence band of ZB type materials is partially degenerated at the $\Gamma-$point (heavy and light hole bands), while in the WZ structure there are three non degenerated valence bands (A, B and C), since we have, additionally to the SO splitting, the crystal field (CF) splitting. Starting from a good description of the valence band in the ZB InAs, we expect to have a realistic band structure for the WZ phase since the different environment of a given atom in the ZB or WZ structures is only at the level of third neighbors. But, in spite of this small difference, the CF splitting is of the order of 100 meV. Based on the quasi-cubic approximation, we obtained $\Delta_{CF}$ and the spin-orbit splitting $\Delta_{SO}$ and compared that last one with that of the ZB InAs. The calculated values of the $E_0$ and $E_1$ gap of WZ InAs were compared with recent experimental results \cite{Moeller2012}. Finally, we analyzed the differences in the dielectric functions of InAs in the ZB and WZ structures and compared them with recent calculations of A. De and C. E. Pryor \cite{Amrit2012}, based on empirical pseudopotentials.

\section{Methods}

We have performed the \textit{ab initio} calculations using the \textit{full potential linearized augmented plane wave method} as implemented in the WIEN2k code \cite{WIEN}. In our calculations, we expanded the basis functions up to $R_{mt} \ast K_{max}=9$, where $R_{mt}$ is the smallest of all atomic sphere radii and $K_{max}$ is the plane wave cut-off. In $k-$space, we used a grid of 16 nonequivalent points in the irreducible part of the Brillouin Zone (a distinct $k-$grid was used only for the dielectric function calculation, see below). For the structural optimization, we neglected the SO interaction and employed the Perdew and Wang (PW) correlation and exchange potential in the \textit{local density approximation} (LDA) \cite{LDA}. It is known that, in most of the cases, LDA gives better structural parameters than the \textit{generalized gradient approximation} (GGA) when compared to experimental results \cite{EngelVosko}. The optimized InAs structure in the ZB phase gives us a lattice constant of $a_{ZB}=6.0407$ \AA, with a deviation smaller than $0.3\%$ from tabulated values \cite{Vurgaftman2001}. Finally, we obtained the optimal structural parameters for WZ InAs and, for the band structure calculation, we replace the PW exchange-correlation potential by the mBJ \cite{Tran2009}. It is also important to emphasize that the SO interaction was fully taken into account in all calculations, except, as mentioned before, in the structural optimization, since it has a negligible contribution to the lattice parameter determination.

We have modeled the WZ InAs NWs using a bulk system, what is easily justified by the large diameters of real NWs. They are usually in the 50-200 nm diameter range and have several micrometers of length \cite{Sun2010,Sun2012,Moeller2012}, what means that carriers confinement and surface effects can be neglected \cite{Cantarero2013}.

\section{Results}

\subsection{Structural Parameters}

For the structural optimization of WZ InAs, we started from the tabulated lattice parameter for ZB InAs \cite{Vurgaftman2001} and obtained the correspondent parameters for the compact WZ structure through $a_{WZ} = a_{ZB}/\sqrt{2}$, $c = \sqrt{8/3}\, a_{WZ} $ and the internal parameter $u=3/8$. After that, we built 25 unit cells with different $c$ parameters and cell volumes ranging from $98\%$ to $102\%$ of these initial values, calculated the total energy for all these cells and obtained the $c$ and $a_{WZ}$ parameters as that corresponding to the minimum total energy through a parabolic least square fit \cite{WIEN}. The last step was to optimize $u$ in the final unit cell.
\begin{table}[h]
\caption{\label{exppar} Optimized structural parameters for WZ InAs and deviations from the compact WZ values derived from the calculated ZB structure.}
\begin{center}
\begin{tabular}{@{}lccc@{}}
\br
Parameters & $a_{WZ}$  & $c$  & $u$  \\
\mr
Values &  4.2564 \AA  & 7.0046 \AA & 0.374 \\
Deviation (\%) & $-0.35$ & $0.42$  & $-0.27$\\
\br
\end{tabular}
\end{center}
\end{table}
The structural parameters obtained after the optimization are presented in Table \ref{exppar}, where they are compared with the compact WZ structure obtained from our optimized ZB InAs system. As observed, the deviation from the packed WZ structure is smaller than $0.5\%$, what means a small degree of spontaneous polarization along the growth axis, which actually can be neglected.

In Table \ref{comppar}, a comparison of our structural parameters with previously published results is given. As it can be seen, there is a remarkable agreement with the theoretical values obtained by C. Panse \textit{et al.} \cite{Panse2011} using the projector-augmented wave method, better than 99.78 \%. At the same time, our results agree better than 99.35 \% with the experiment. In particular, the c-axis obtained in the present work agrees very well with the experimental values reported in Ref. \cite{Zanolli2007}, around 99.87\%. The results supplied in Tables 1 and 2 indicate the accuracy of the Wien2k approach.
\begin{table}
\caption{\label{comppar} Our calculated structural parameters for WZ InAs compared with previously published theoretical (theor.) and experimental (exp.) results.}
\begin{center}
\begin{tabular}{@{}lcc@{}}
\br
Parameters & $a_{WZ}$ (\AA)  & $c$ (\AA) \\
\mr
present work &  4.2564 & 7.0046 \\
Ref. \cite{Panse2011} (theor.) & 4.2570 & 6.9894 \\
Ref.  \cite{Zanolli2007} (theor.) & 4.2663 & 6.9669 \\
Ref.  \cite{Kriegner2011} (exp.) & 4.2742 & 7.0250 \\
Ref.  \cite{Zanolli2007} (exp.) & 4.2839 & 6.9954\\
\br
\end{tabular}
\end{center}
\end{table}

\subsection{Band Structure}

In order to obtain an accurate gap for InAs in the WZ phase, and clarify the experimental divergences, we employed the Perdew and Wang LDA correlation plus mBJ exchange potential \cite{Tran2009}. As mentioned before, this last potential has three parameters that can be optimized in order to better reproduce the real interactions \cite{Tran2009,Koller2011,Koller2012}. These values are actually under debate \cite{Koller2012} as previously commented. The mBJ parameter $\alpha$ ($A$ in Ref. \cite{Koller2012}) is given by $\alpha=0.488$ for the general case and $\alpha=0.267$ for semiconductors,
restricted to low gap semiconductors. In the same way, the parameter $\beta=0.500$ Bohr$^{-1/2}$ ($B$ in Ref. \cite{Koller2012}) and $\beta=0.656$ Bohr$^{-1/2}$ were employed for the general case and that of semiconductors, respectively. We have chosen an intermediate value of the parameters $\alpha=0.31$ and $\beta=0.585$ Bohr$^{-1/2}$ to reproduce the well established ZB InAs band gap and the parameter $e=1$ as proposed by K\"{o}ller \textit{et al.} \cite{Koller2012}. This set of parameters give a gap of 0.416 eV for InAs in the ZB phase, in remarkable quantitative agreement with the experimental value \cite{Vurgaftman2001,Moeller2012}. Since the ZB and WZ structures are very similar, the only difference being the packing in the [111] direction ([0001] in the hexagonal lattice), it is reasonable to keep these ZB parameters. This property was exploited in previous works \cite{Muranaka}, where empirical potentials were transferred from ZB to WZ structures assuming that they are not significantly affected by the different atomic stacking \cite{Pryor2010}, although the CF splitting is not negligible. Using the parameters given above, we obtain a band gap of 0.461 eV for the WZ phase, in perfect quantitative agreement with previous experimental results \cite{Moeller2012}. It is also important to say that the value obtained for the $E_1$ gap (2.25 eV)  agrees very well with the experiment too \cite{Moeller2011,Zardo2013}.

\begin{figure}[htb]
\begin{center}
\includegraphics[width=0.4\textwidth]{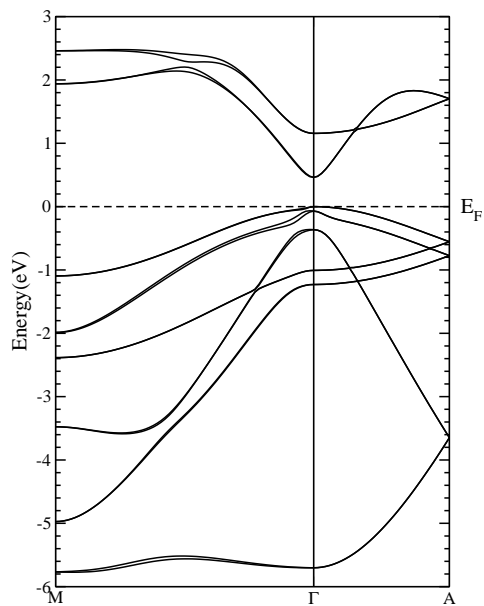}
\caption{\label{bands} Band structure for InAs in the WZ phase, calculated using the mBJ+LDA potential, in the $M-\Gamma-A$ directions of the Brillouin zone.}
\end{center}
\end{figure}

In Fig. \ref{bands}, we show the calculated band structure for WZ InAs in the $M-\Gamma-A$ directions. Labeling the three uppermost valence bands as $A$, $B$ and $C$ in decreasing energy order, the obtained splitting among them at the $\Gamma$ point are $\Delta_{A-B} = 0.070$ eV and $\Delta_{A-C} = 0.364$ eV. Using the quasi-cubic approximation, the values obtained for $\Delta_{SO}$ and $\Delta_{CF}$ are 0.312 and 0.122 eV, respectively. The first value is basically the same as that obtained for the ZB phase in our calculations (the experimental value is $\approx 0.4$ eV, of the same order as the gap). If we compare $\Delta_{SO}$ of WZ InAs with that of the WZ InP, 0.084 eV  \cite{InPwurtSSC}, it is significantly larger than in the present calculations. The reason is that As is heavier than P and thus relativistic effects are more important in InAs. On the other hand, the calculated $\Delta_{CF}$ is comparable to that of WZ InP, 0.147 eV \cite{InPwurtSSC}, which makes sense since the crystal field must be similar in the two compounds.

\begin{figure}[htb]
\begin{center}
\includegraphics[width=0.4\textwidth]{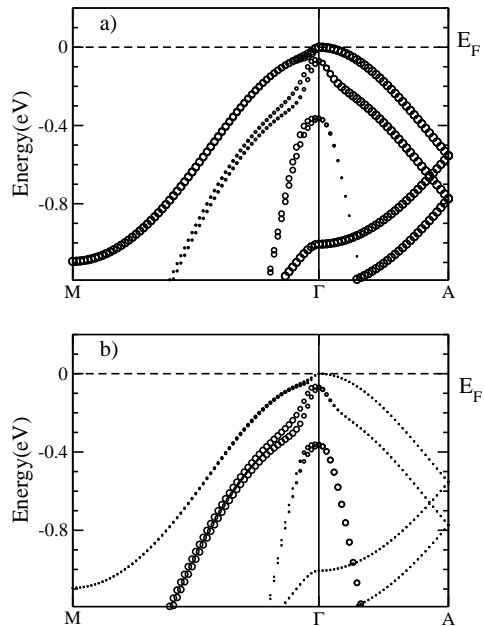}
\caption{\label{char} Band character of InAs in the WZ phase [(a) $p_xp_y$ orbital and (b) $p_z$ orbital]. The size of the circles is proportional to the band character.}
\end{center}
\end{figure}

The band character of the top valence bands is presented in Fig. \ref{char}. In this plot, the specific orbital contribution to the corresponding electron wave function (band character) is shown. The $p_xp_y$ character is presented in panel (a), while the $p_z$ character is shown in panel (b) (the $z$-axis is equivalent to the WZ $c$-axis). In both cases, the circles size is proportional to the assigned character. Remembering that the first conduction band is mainly $s-$type, our results show that the transitions between the $A-$valence band and the lowest conduction band will be allowed for in-plane ($xy$) polarization, no matter you consider the $M-\Gamma$ or the $\Gamma-A$ direction in the Brillouin zone. A different behavior takes place for the $B-$valence bands, which has a strong $p_z$ character in $M-\Gamma$ direction, but a $p_xp_y$ character in $\Gamma-A$. This behavior is inverted for the $C-$valence band.

\begin{figure}[htb]
\begin{center}
\includegraphics[width=0.4\textwidth]{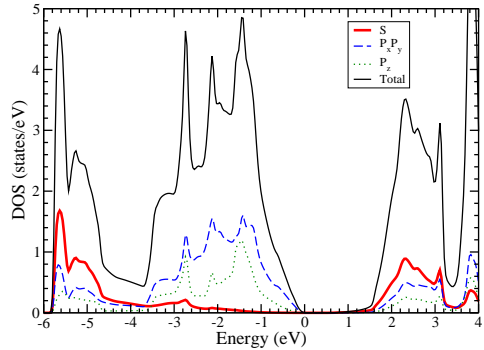}
\caption{\label{DOS} Total and partial density of states for InAs in the WZ phase: $s$-type (thick solid line $-$red online$-$), $p_xp_y$-type (dashed line $-$blue online$-$), $p_z$-type (dotted line $-$green online$-$) and total density of states (black solid line).}
\end{center}
\end{figure}

Another way to quantify the strength the matrix elements for inter-band transitions is through the analysis of the density of states (DOS). It is important to remark that the projected DOS curves are calculated only within the muffin-tin region, the small contribution of the interstitial region is taken into account only in the calculation of the total DOS. In Figure \ref{DOS}, we show the results for the $s$-projection (thick solid line $-$red online$-$), $p_xp_y$-projection (dashed line $-$blue online$-$), $p_z$-projection (dotted line $-$green online$-$) and the total DOS (black solid line). The $d$-orbital contribution is significant only below $-14$ eV and it has been omitted. As one can see, the bottom of the conduction band is dominated by $s$-type states, while the main contribution to the top of the valence band is given by $p_xp_y$-type states with a minor, but not negligible $p_z$ character. In a DOS plot the partial contribution of the different orbitals are clearly observed, but the selection rules for optical transitions are better analyzed in a band structure plot with character, as that shown in Fig. \ref{char}.

\subsection{Dielectric Function}

\begin{figure}[htb]
\begin{center}
\includegraphics[width=0.5\textheight]{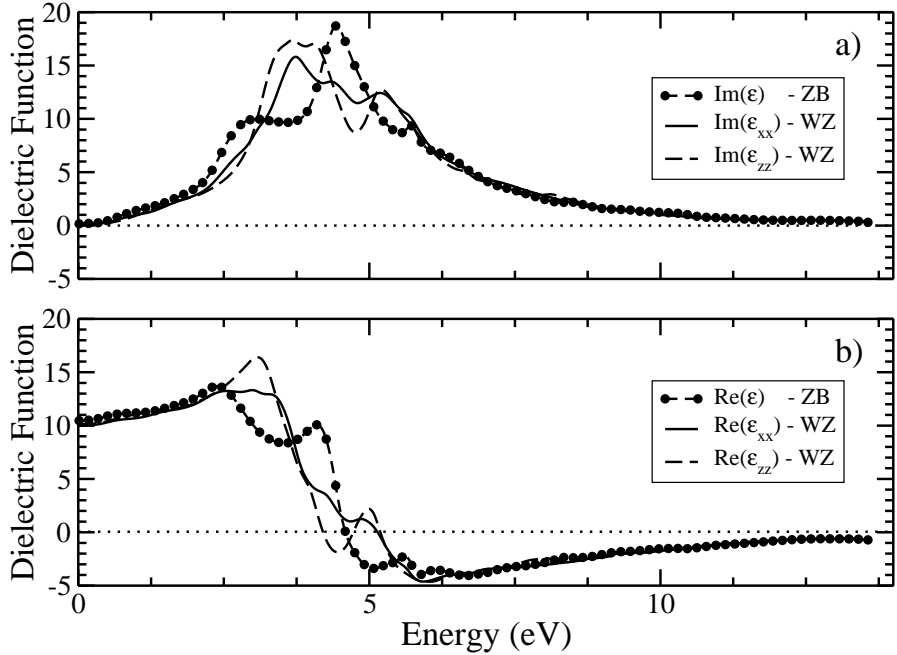}
\caption{\label{dielectric} Imaginary (a) and real (b) parts of the complex dielectric function of WZ InAs as a function of incident photon energy for polarizations in the $xy$ plane and along the $z$-axis. The results for ZB InAs are included for comparison.}
\end{center}
\end{figure}

We have calculated the ZB and WZ InAs dielectric functions using the OPTIC facility provided within the WIEN2k package \cite{optic}. This calculation requires a dense $k$-mesh in the Brillouin zone and, in our case, convergence was obtained with 1200 $k$-points (84 nonequivalent). The results are shown in Fig. \ref{dielectric}.

First of all, the dielectric function for the ZB structure is very similar to that measured by ellipsometry \cite{Adachi1987}. In the imaginary part of the dielectric function there are two main peaks, one close to 3 eV, and a second peak around 4.6 eV, corresponding to the $E_1$ and $E_2$ electronic transitions. The absolute value of the calculated dielectric function is of the same order (around 20) than the experimental one, although the first peak seems to be underestimate. This fact, and the extension of the imaginary part of $\varepsilon$ below the band gap could be related with the broadening of 0.1 eV included to avoid the numerical noise produced by the finite grid. In the measured Re$(\varepsilon)$ there there are also two peaks, as in the experiment \cite{Adachi1987}, although the first peak has a larger value of the dielectric function, just the opposite as the observed in the experiment \cite{Adachi1987}.

In the WZ InAs case, the dielectric function in the region of optical absorption has a richer structure when compared to that of the ZB structure. There is, additionally, the expected anisotropy, giving a higher variation for $\varepsilon_{zz}$ than for $\varepsilon_{xx}$ for both the real and imaginary parts. This anisotropy can be related to the extra number of bands (around the $E_1$ critical point for instance) due to the lack of degeneracy in the valence band or, in other words, to the crystal field splitting. Comparing our results with other calculations, one can say that there is a good qualitative agreement, although in our opinion there is an overestimation in the real part of the dielectric function in Ref. \cite{Amrit2012}, with a value close to 40. In our case, the dielectric function is very similar to that of the ZB phase, the only difference being the anisotropy.

\section{Conclusions}

We have presented \textit{ab initio} calculations of the structural, electronic and optical properties of WZ InAs. Despite being a technologically important material, there are still open questions related to its basic properties, the most significant one being its band gap value. To clarify them, we employed the LDA correlation plus the mBJ exchange potentials through the \textit{full potential linearized augmented plane wave method} as implemented in the WIEN2k code. In a first step, we obtained the mBJ parameters that best reproduce the electronic interactions in the ZB phase of InAs and then, taking advantage of ZB and WZ similarities, we transferred them to the WZ system and obtained a band gap value of 0.461 eV, in perfect agreement with recent experimental results \cite{Moeller2012}. The value obtained for the $E_1$ gap, 2.25 eV, also agrees very well with the experiment \cite{Moeller2011a}. The spin orbit splitting is the same in the ZB and WZ phases, and there is a small CF splitting of 0.122 eV, due to the anisotropy of the hexagonal lattice. Remembering that the band gap value is one of the most difficult quantities to calculate, we provided electronic and optical properties of WZ InAs with a high degree of confidence. Our results show that the WZ InAs lattice has a small deviation from the  ``ideal'' (packed) WZ structure (less than $0.5\%$) and the complex dielectric function calculated show smoother curves with more reasonable values than previously published theoretical results \cite{Amrit2012}. We expect that this work can stimulate more experimental studies on WZ InAs.

\section{Acknowledgments}

This work was partially supported by the projects BEX1042/08-4 and 133/2007 (CAPES-MECD-DGU, Brazil) and MAT2012-33483 and CDS2010-0044 (Ministry of Finances and Competitiveness of Spain). Part of calculations were done with the supercomputers TIRANT and Lluis Vives, from the University of Valencia.

\section*{References}
\providecommand{\newblock}{}

%\bibliography{inas-SST}
\end{document}